\documentclass[sigconf, nonacm]{acmart}

\newcommand\vldbyear{2026}
\newcommand\vldbworkshop{Applied AI for Database Systems and Applications (AIDB 2026)}
\newcommand\vldbauthors{\authors}
\newcommand\vldbtitle{Hollywood: Towards a Large Movie Dataset for Database Benchmarking}
\newcommand\vldbavailabilityurl{https://github.com/utndatasystems/hollywood}
\newcommand\vldbpagestyle{empty}

\usepackage{enumitem}
\setlist[itemize]{leftmargin=24pt}

\usepackage{booktabs}
\usepackage{graphicx}
\usepackage{array}
\usepackage{colortbl}
\usepackage{amsmath}
\usepackage{siunitx}
\usepackage{xspace}
\usepackage{caption}
\usepackage{subcaption}
\usepackage{url}
\usepackage{placeins}
\makeatletter
\@ifundefined{@makespecialcolbox}{%
  \@ifundefined{@make@specialcolbox}{}{%
    \let\@makespecialcolbox\@make@specialcolbox
  }%
}{}
\makeatother
\usepackage{cuted}
\usepackage{dblfloatfix}

\graphicspath{{figures_main/}}
\definecolor{blockTwo}{HTML}{0B7D8B}
\definecolor{blockThree}{HTML}{168A1B}
\definecolor{blockFour}{HTML}{B83220}
\newcommand{\figblock}[2]{\textcolor{#1}{\textcircled{\raisebox{-0.35pt}{\small\bfseries #2}}}}
\newcommand{\blocktwo}{\figblock{blockTwo}{2}\,}
\newcommand{\blockthree}{\figblock{blockThree}{3}\,}
\newcommand{\blockfour}{\figblock{blockFour}{4}\,}

\newcommand{\system}{Hollywood\xspace}
\newcommand{\qerr}{Q-error\xspace}
\newcommand{\pg}{PostgreSQL\xspace}
\newcommand{\duck}{DuckDB\xspace}
\newcommand{\mscn}{MSCN\xspace}
\newcommand{\zeroshot}{ZeroShot\xspace}

\newcommand{\sparagraph}[1]{\vspace{1mm}\noindent {\bf #1}}
\newcommand{\topic}[1]{\sparagraph{#1.}}

\title{Hollywood: Towards a Large Movie Dataset for Database Benchmarking}

\author{Ivan Iachnyk}
\affiliation{\institution{University of Technology Nuremberg}\city{Nuremberg}\country{Germany}}
\email{ivan.iachnyk@utn.de}

\author{Mihail Stoian}
\affiliation{\institution{University of Technology Nuremberg}\city{Nuremberg}\country{Germany}}
\email{mihail.stoian@utn.de}

\author{Andreas Kipf}
\affiliation{\institution{University of Technology Nuremberg}\city{Nuremberg}\country{Germany}}
\email{andreas.kipf@utn.de}

\begin{document}

\begin{abstract}
The IMDb real-world dataset of the JOB benchmark has been extensively used in the last decade as part of the research line on cardinality estimation, given its ability to stress test both traditional and learned estimators. However, unlike the synthetic TPC family, it does not come with a scale factor, being a simple dump.

We introduce \system, a synthetic IMDb-compatible benchmark generator that combines LLM-generated semantic dictionaries with deterministic temporal-graph-based relational data generation. We analyze a preliminary Hollywood-200K, which contains \num{200000} primary movies, generated series and episode title rows, 19.7M IMDb-style rows, and 213 nonzero JOB-Light, JOB, and JOB-Complex queries. Experiments with two open systems demonstrate that Hollywood induces cardinality estimation errors comparable to or exceeding those observed on the original IMDb dataset. The release includes generation settings and prompt/LLM-output provenance together with adapted SQL and labels, enabling tests of whether cardinality estimators generalize beyond a fixed movie snapshot and distribution.
\end{abstract}

\maketitle

\pagestyle{\vldbpagestyle}
\begingroup\small\noindent\raggedright\textbf{VLDB Workshop Reference Format:}\\
\vldbauthors. \vldbtitle. VLDB \vldbyear\ Workshop: \vldbworkshop.\\
\endgroup
\begingroup
\renewcommand\thefootnote{}\footnote{\noindent
This work is licensed under the Creative Commons BY-NC-ND 4.0 International License. Visit \url{https://creativecommons.org/licenses/by-nc-nd/4.0/} to view a copy of this license. For any use beyond those covered by this license, obtain permission by emailing \href{mailto:info@vldb.org}{info@vldb.org}. Copyright is held by the owner/author(s). Publication rights licensed to the VLDB Endowment. \\
\raggedright Proceedings of the VLDB Endowment.
ISSN 2150-8097. \\
}\addtocounter{footnote}{-1}\endgroup

\ifdefempty{\vldbavailabilityurl}{}{
\vspace{.3cm}
\begingroup\small\noindent\raggedright\textbf{VLDB Workshop Artifact Availability:}\\
The source code, data, and/or other artifacts have been made available at \url{\vldbavailabilityurl}.
\endgroup
}

\section{Introduction}

\topic{Motivation}
Query optimizers are commonly evaluated on benchmarks that make different compromises. TPC-H, TPC-DS, and SSB provide controlled scale factors and reproducible data generation~\cite{tpch,tpcds,oneil2009ssb}, following a long line of synthetic database generators~\cite{gray1994billion,bruno2005flexible,lo2014mybenchmark,ding2021dsb}. By contrast, the JOB benchmark family~\cite{leis2015job, kipf2019mscn, wehrstein2025jobcomplex} uses a 2013 snapshot of IMDb, exposing optimizers to string-heavy predicates and correlated multi-join queries~\cite{leis2015job}. Despite its small size, it has been \emph{the} dataset for learned-cardinality, learned-cost, and learned-optimizer work~\cite{kipf2019mscn,hilprecht2020deepdb,hilprecht2022zeroshot,marcus2019neo,marcus2021bao}.

\topic{The Gap}
The cost of JOB's realism is that the data instance is fixed, making it difficult to ask whether an estimator generalizes to a structurally similar but previously unseen data distribution. Recent work revisits JOB~\cite{leis2025stillasking}, extends JOB-style query families~\cite{wehrstein2025jobcomplex} or synthesizes trace-shaped cloud workloads over existing datasets, including the IMDb dump~\cite{DBLP:journals/corr/abs-2506-12488, wehrstein2025redbench, schmidt2025sqlstorm}. Other approaches generate relational data using query-aware or constraint-driven methods~\cite{binnig2007qagen,arasu2011declarative,sanghi2018hydra,sanghi2021hifi,sanghi2022projection,sanghi2022joinconstraints}, or graph-conditional diffusion~\cite{ketata2025grdm}, while GReaT~\cite{DBLP:conf/iclr/BorisovSLPK23}, SPADA~\cite{DBLP:conf/emnlp/YangZPK25}, and GraDe~\cite{DBLP:conf/emnlp/ZhangYPK25} use LLMs to capture dependencies among attributes of a single relation. However, all three methods are defined for a single table and do not generate database instances spanning multiple relations. The IMDb/JOB ecosystem therefore still lacks a generator that retains JOB's correlated predicates.

\topic{Hollywood} In this work, we address this missing path. Given a generation profile specifying primary-movie count, years, entity-pool sizes, and a seed, Hollywood produces a fresh IMDb-style database with jointly generated movie, person, company, and literal values. The LLM component runs through Google's Gemini API~\cite{google2026gemini31flashlite}; in the current prompting profile, movie-plot generation costs roughly US\$100 per \num{100000} primary movies, excluding optional series/episode plot text. For Hollywood-200K, we additionally adapt JOB-family templates to existing literals, execute them to obtain labels, and release the files needed to reproduce the learned-estimator measurements on the released database.

We focus on Hollywood-200K, a 200,000-primary-movie export with 351,455 IMDb title rows, and evaluate adapted JOB-Light, JOB, and JOB-Complex query workloads on it. We compare it with two IMDb references: the original benchmark workload on the full IMDb dataset and a matched sample with the same title-row count.

\topic{Contributions}
Thus, our contributions are:
\begin{itemize}

\item[(1)] an IMDb-compatible generator combining LLM semantic priors, deterministic temporal-graph materialization, and strict relational export,
\item[(2)] an evaluation of \pg, \duck, \mscn, and \zeroshot on Hollywood, with full-query, single-table selection, and selected-plan cost references against IMDb references,
\item[(3)] and a release package containing the generator, the Hollywood-200K export, the nonzero adapted JOB-Light, JOB, and JOB-Complex workloads used here, and the evaluation artifacts needed to reproduce the reported measurements.
\end{itemize}
\topic{Outlook} Building on the framework presented here, we plan to release Hollywood at scale factors 10 and 100, treating IMDb as sf=1, and replay all existing evaluations of learned cardinality estimators beyond the widely used, yet relatively small, IMDb dataset. We next outline Hollywood's data generation process.

\begin{figure*}[t!]
  \centering
  \includegraphics[width=0.86\textwidth]{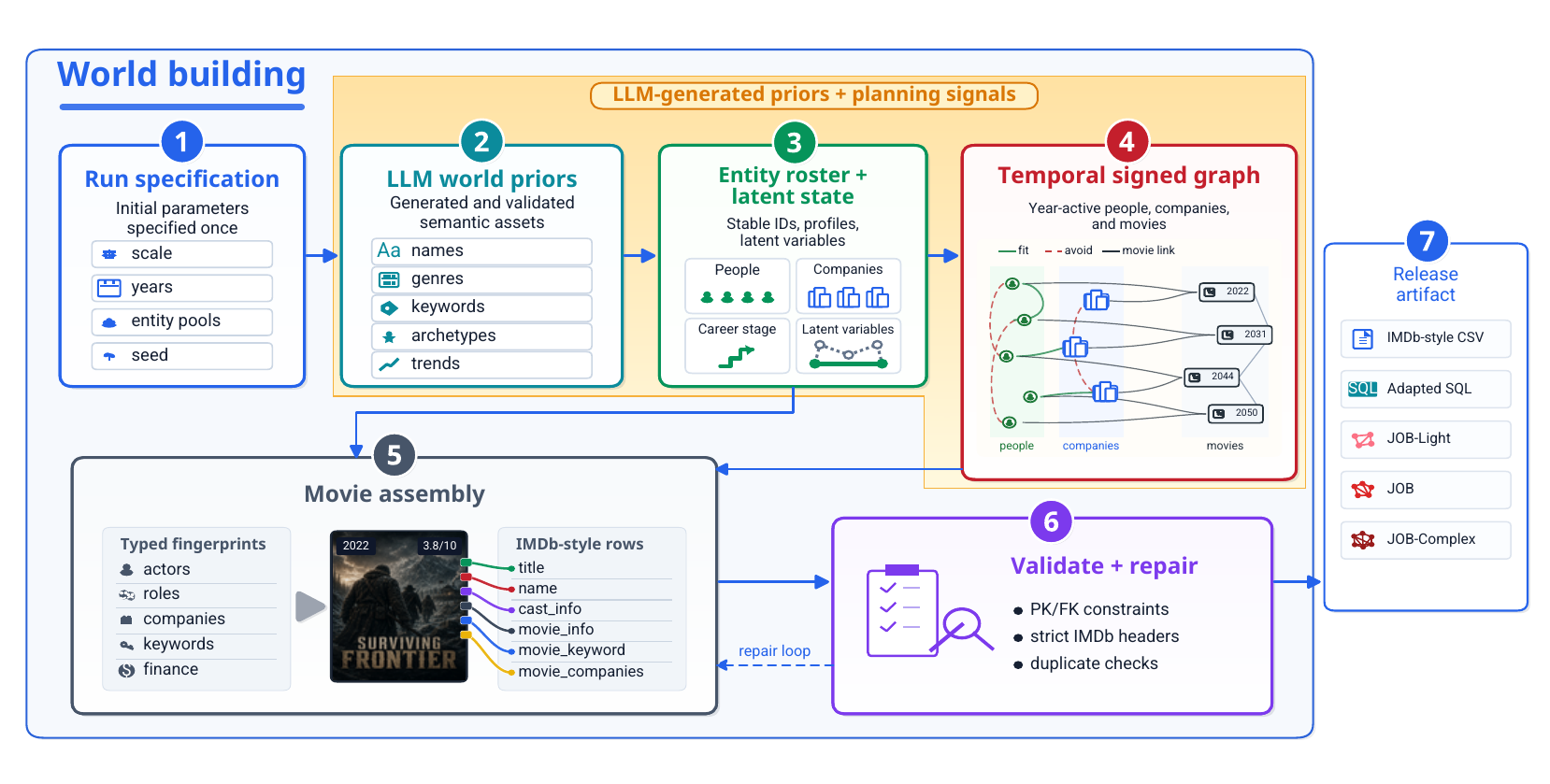}
  \caption{Hollywood generation flow from LLM-authored priors to temporal graph state, movie assembly, validation/repair, and the released IMDb-style CSV plus adapted SQL artifact.}
  \Description{The Hollywood pipeline has seven blocks: run specification, LLM world priors, entity roster and latent state, temporal signed graph, movie assembly into IMDb-style rows, validation and repair with a feedback loop, and a release artifact containing IMDb-style CSV files and adapted JOB-family SQL.}
  \label{fig:system}
\end{figure*}

\section{Generation Method}

\topic{Construction Flow}
Fig.~\ref{fig:system} sketches the construction path. A run specification fixes scale, years, entity pools, and seed, after which LLM priors, entity rosters, a temporal signed graph, and movie assembly produce a validated IMDb-style export. The evaluated artifact packages the export with adapted SQL workloads, labels, plans, runtime records, and validation and repair reports, making the benchmark instance inspectable and replayable.

\topic{Export Scale}
Our current IMDb/JOB export contains \num{200000} primary movies plus generated TV-series and episode title rows, yielding \num{351455} title rows, \num{480000} name rows, 4.98M cast rows, 2.10M movie-keyword rows, 5.39M movie-info rows, and 19.7M rows overall. The matched IMDb sample has the same title-row count and a similar largest JOB-facing relationship table: 5.25M \texttt{cast\_info} rows versus Hollywood's 4.98M. It is therefore a title/cast-scale reference, not a distributional twin.

\topic{Rows, SQL, and Labels}
The generator holds typed profiles for people and movies. For a movie $m$ generated in year $y$, its profile $z_m$ and the temporal graph snapshot $G_y$ active in that year jointly condition the selection of cast slots, companies, keywords, and metadata. The exporter then emits coordinated rows in \texttt{name}, \texttt{title}, \texttt{cast\_info}, \texttt{movie\_info}, \texttt{movie\_companies}, and \texttt{movie\_keyword}. Thus predicates over names, genres, companies, keywords, and cast roles originate from one generated movie rather than independent per-table draws. The release also includes the final adapted SQL workloads. Adaptation performs a bounded search over replacements for hard-coded literals using values found in the Hollywood-200K database, while preserving aliases, joins, operators, grouping, and count wrappers. It retains candidates that parse, execute, and have positive labels, favoring those that stress both \pg and \duck; the database remains fixed throughout this search.

\begin{figure*}
\centering
\includegraphics[width=0.96\textwidth]{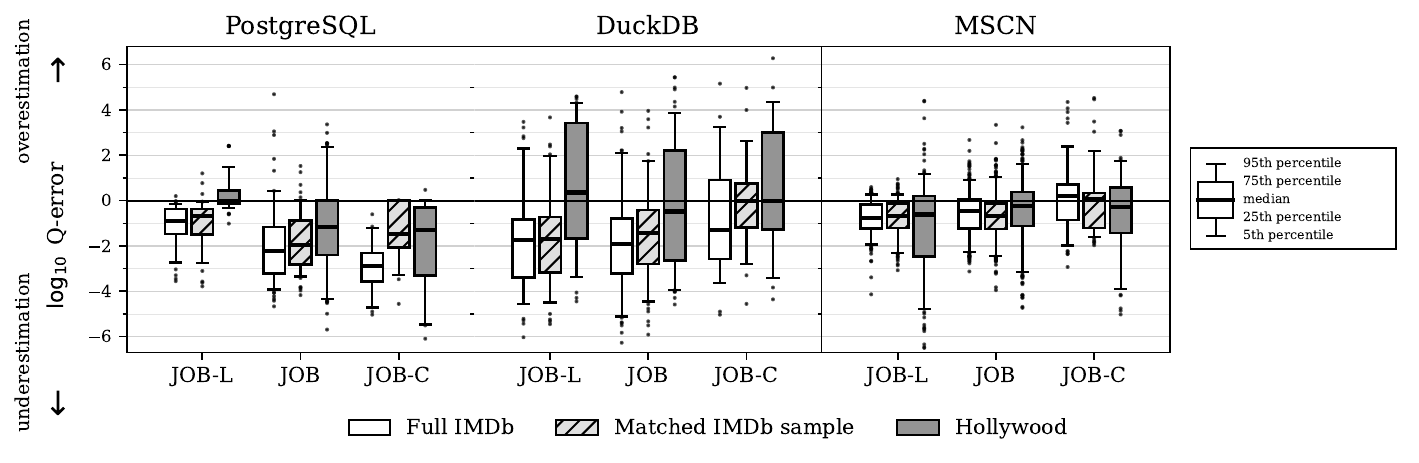}
\caption{Signed full-query cardinality error. Boxes compare full IMDb, the matched IMDb sample, and Hollywood; \mscn uses separately trained models for each dataset.}
\Description{Three panels of signed full-query cardinality boxplots for PostgreSQL, DuckDB, and MSCN. Each panel compares full IMDb, a matched IMDb sample, and Hollywood across JOB-Light, JOB, and JOB-Complex.}
\label{fig:full-query-signed}
\end{figure*}

\subsection{LLM Priors}

\topic{Semantic Priors}
\system uses LLMs as semantic prior generators, not tuple emitters. In \blocktwo of Fig.~\ref{fig:system}, LLM calls use Gemini 3.1 Flash-Lite Preview~\cite{google2026gemini31flashlite}; the exact API model identifier is recorded in artifact provenance. Asking an LLM to write millions of rows would be expensive, hard to audit, and difficult to repair, especially because long-context studies show that larger windows do not guarantee reliable use of every relevant fact~\cite{liu2024lostmiddle,hsieh2024ruler}. Instead, bounded prompts produce reusable lexical assets, entity priors, and temporal priors that are parsed as JSON, normalized, deduplicated, checked for coverage, versioned, and consumed by seeded code.

\topic{World-Building Bias}
LLM-generated latents in \blockthree provide one input to later deterministic selectors. A purely random latent table would not automatically couple names, regions, genres, career stages, company strategies, and risk profiles. We therefore use the model as a nonuniform world-building prior: its co-occurrence patterns couple otherwise independent labels, but training-data biases can also enter the generated priors~\cite{bender2021stochasticparrots,bommasani2021foundation}. The artifact stores these versioned priors separately from tuple materialization. Seeded selectors combine them with graph state and validation rules when materializing rows.

\subsection{Temporal Graph}

\topic{Graph State}
The temporal graph in \blockfour stores cross-entity dependencies that are reused when movies are assembled. Its nodes represent people and companies; movies are assembled using the graph but are not themselves graph nodes. Each edge has a type, sign, strength, provenance record, and active-year interval. Positive edges encode collaboration, mentorship, genre fit, and company--talent affinity, while negative edges encode rivalry, avoid lists, brand mismatch, or market competition. Multiple selectors reuse these signed edges during movie assembly, so a casting choice can also affect company links, keywords, and metadata rather than leaving each relation group to be sampled separately.

\topic{Graph Sampling}
Initial edges are stochastic but feature-driven. Candidate pairs come from genre, market, community, agency, career-stage, and company pools, then weighted draws combine creative-style similarity, genre overlap, risk compatibility, prior workload, degree caps, and seeded noise. Company-person and company-company edges use analogous genre, budget, market, risk, and strategy features. The generator consumes the graph through an active-at-year view: assembly asks for friendship, rivalry, director-preference, avoid, and company-affinity edges active in the target year. Movie assembly and year-boundary evolution can add, update, or expire edges before later years are generated. The history borrows slowly changing dimension type-2 semantics~\cite{kimball2013datawarehouse}: updates close the old active interval and insert a replacement edge version, while expirations only close the interval, preserving which edge version was active in each synthetic year.

\subsection{Chronological Assembly}

\topic{Yearly Assembly}
\system first samples a yearly slate of movie slots. It then materializes each movie in chronological order by selecting concept, title, people, companies, keywords, and metadata. For a movie profile $z_m$ in year $y$, let $x$ be a candidate person, company, or keyword and let $G_y$ be the graph snapshot active in $y$. Its selection weight is schematically
\[
\begin{aligned}
\Pr(x\mid z_m,G_y,y)\propto{}& w_{\mathrm{pop}}(x,y)\cdot w_{\mathrm{fit}}(x,z_m)\cdot w_{\mathrm{cap}}(x,y)\\
&\cdot w_{\mathrm{latent}}(x,z_m)\cdot w_{\mathrm{graph}}(x,G_y)\cdot w_{\mathrm{policy}}(x,y).
\end{aligned}
\]
The factors represent current popularity or activity, movie-role compatibility, remaining yearly capacity, latent-profile similarity, affinity in the active graph, and policy constraints such as diversity or degree caps. Each selector uses the applicable factors and caps for its candidate type. After each year, coappearances, company activity, genre movement, and performance signals are summarized into graph operations, latent deltas, stage/tier updates, retirements, dissolutions, and genre shifts. These changes condition the next year, making the graph dynamic and probabilistic while preserving chronological consistency.

\section{Evaluation}

The evaluation separates base-selection difficulty from full-query cardinality and selected-plan runtime prediction. The comparison scope depends on the task: full-query cardinality compares Hollywood with both full IMDb and a title-count-matched IMDb sample, single-table selections isolate local predicate difficulty on the matched setting, and selected-plan runtime prediction compares Hollywood with the matched IMDb sample on PostgreSQL plans.

\subsection{Setup}

\topic{Workloads and References}
All measurements use the frozen Hollywood-200K database and the released adapted SQL and label files. The released instance includes plot rows for movies but not for TV series or episodes. The Hollywood workload contains 70 JOB-Light, 113 JOB, and 30 JOB-Complex queries selected from adapted candidates with positive labels~\cite{leis2015job,wehrstein2025jobcomplex}. As references, full IMDb uses the canonical benchmark workload, while the matched IMDb sample uses adapted literals at the same title-row scale.\footnote{Note that 28 of the 30 literal-adapted JOB-Complex queries on the matched IMDb sample have nonzero cardinality; all full-query summaries use this fixed subset.} Single-table selection labels are exact filtered base-relation counts obtained with the canonical \pg IMDb schema.

\begin{figure}[t]
  \centering
  \includegraphics[width=\columnwidth]{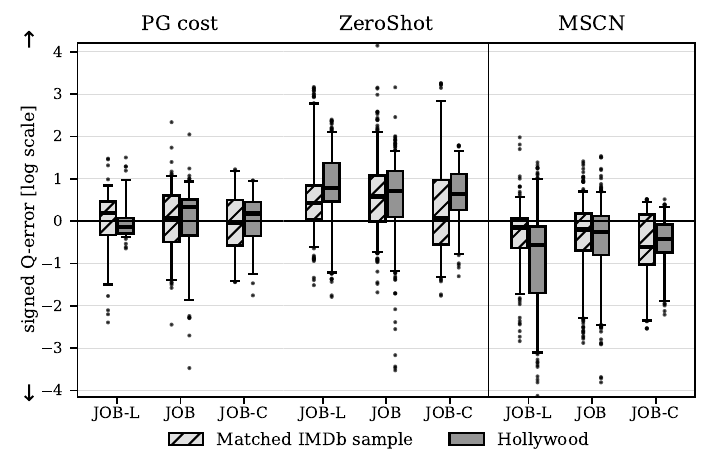}
  \caption{\raggedright Signed selected-plan runtime-prediction error comparing matched IMDb and Hollywood. PG cost denotes the PostgreSQL optimizer cost. Positive values indicate overprediction and negative values indicate underprediction.}
\Description{Boxplots comparing signed selected-plan runtime prediction errors for PostgreSQL cost, ZeroShot, and MSCN cost on JOB-Light, JOB, and JOB-Complex. Each estimator compares the matched IMDb sample with Hollywood.}
  \label{fig:cost}
\end{figure}

\topic{Estimators}
We use \pg 16.14~\cite{postgresql2026docs} and \duck 1.5~\cite{raasveldt2019duckdb} as traditional cardinality estimators. For learned models, \mscn~\cite{kipf2019mscn} is evaluated for cardinality and selected-plan cost, while \zeroshot~\cite{hilprecht2022zeroshot} is evaluated for selected-plan cost; \pg cost provides a conventional selected-plan runtime proxy. \pg and \duck run on explicit IMDb-compatible schemas; full-query cardinalities use \pg planner rows and the first \duck cardinality marker below count-style aggregates. The MSCN cardinality model follows SetConv with table, predicate, join, and sample-bitmap inputs, plus our deterministic string-hash encoding for string literals. For \mscn, cardinality models train separately per plotted dataset (90K full-query examples, 100 epochs, repeated seeds); selected-plan cost models do the same for Fig.~\ref{fig:cost} datasets with 10K plan examples.

\topic{Runtime Prediction}
For each query, PostgreSQL selects a physical plan whose measured execution time is the prediction target. We compare \pg cost, \zeroshot, and an MSCN cost model on that plan, with intra-query parallelism disabled because gather and parallel-aware operators are outside the learned-plan vocabulary. We map \pg cost to milliseconds with a workload-specific geometric-mean scale. \zeroshot uses the pretrained cross-database cost model of Hilprecht et al.~\cite{hilprecht2022zeroshot} under the evaluation setup of Heinrich et al.~\cite{heinrich2025lcm}.

\subsection{Full-Query Cardinality}

\topic{PostgreSQL}
For full-query cardinality, Fig.~\ref{fig:full-query-signed} compares Hollywood with two IMDb references: the canonical JOB-family workload on full IMDb and a literal-rebound workload on a same-title-count IMDb sample. \pg has smaller Hollywood medians on JOB and JOB-Complex than full IMDb (\num{28.0} vs. \num{261}, and \num{20.5} vs. \num{758}), but the tail is larger: p95 \qerr reaches \num{2.37e4} on JOB and \num{3.07e5} on JOB-Complex, above the full-IMDb p95 values of \num{1.15e4} and \num{6.64e4}. The matched IMDb sample shows the same contrast, with lower Hollywood medians but p95 values about \num{9.9} times larger on JOB and \num{118} times larger on JOB-Complex. In the signed boxes, these large \pg errors mostly lie below zero; together with the much smaller base-selection errors in Table~\ref{tab:base}, this pattern is consistent with cardinality underestimation emerging after relationship joins.

\topic{DuckDB}
\duck gives the contrasting case. Hollywood shifts its signed boxes upward compared with the IMDb references: on JOB-Complex, median \qerr is \num{190}, compared with \num{117} on full IMDb and \num{11.7} on the matched IMDb sample; on JOB-Light, median signed log error moves from about \num{-1.7} on both IMDb references to \num{0.36} on Hollywood. The same export therefore produces predominantly negative post-join errors for \pg and frequent overestimation for \duck.

\topic{Learned Estimator}
For JOB-Light, JOB, and JOB-Complex, \mscn medians remain low: full IMDb (\num{5.89}, \num{4.10}, \num{6.62}), matched IMDb (\num{5.02}, \num{5.81}, \num{5.50}), and Hollywood (\num{7.47}, \num{4.23}, \num{5.32}). Tails differ more: Hollywood p95s (\num{7.12e4}, \num{1480}, \num{1.14e4}) exceed full IMDb (\num{90.1}, \num{194}, \num{1874}) and matched IMDb (\num{210}, \num{298}, \num{155}), so typical learned errors are small but Hollywood contributes large tail cases.

\subsection{Base Predicates}

Tab.~\ref{tab:base} separates local predicate difficulty from join difficulty: for \pg, Hollywood single-table selection p95 is only \num{2.30}, \num{22.0}, and \num{48.0} on JOB-Light, JOB, and JOB-Complex, while full-query p95 reaches \num{2.37e4} on JOB and \num{3.07e5} on JOB-Complex. These results are consistent with the largest \pg errors emerging after relationship joins rather than in individual filtered base-relation estimates, while \duck's largest single-table deviations are a few JOB-Light \texttt{movie\_info\_idx} rating outliers.

\begin{table}[h]
\centering
\caption{Single-table selection \qerr. Cells report median / p95 for filtered base-relation predicates; Hollywood columns are shaded.}
\label{tab:base}
\small

\begin{tabular}{@{}llc>{\columncolor{black!6}}c@{}}
\toprule
Workload & Estimator & Matched IMDb sample & Hollywood \\
\midrule
JOB-L & PG & \num{1.01} / \num{1.85} & \num{1.01} / \num{2.30} \\
 & DuckDB & \num{2.28} / \num{91.6} & \num{4.86} / \num{3.52e4} \\
JOB & PG & \num{1.00} / \num{14.2} & \num{1.00} / \num{22.0} \\
 & DuckDB & \num{2.40} / \num{4830} & \num{2.23} / \num{3600} \\
JOB-C & PG & \num{1.00} / \num{9.98} & \num{1.00} / \num{48.0} \\
 & DuckDB & \num{2.01} / \num{2328} & \num{2.28} / \num{1248} \\
\bottomrule
\end{tabular}

\end{table}

\subsection{Selected-Plan Runtime Prediction}

\topic{Selected PostgreSQL Plans}
Following learned-cost-model evaluations on JOB-derived tasks~\cite{heinrich2025lcm}, Fig.~\ref{fig:cost} evaluates runtime prediction on fixed selected PostgreSQL plans. With the workload-specific scale, \pg cost has similar p95 error on Hollywood and matched IMDb JOB-Complex (\num{24.2} vs. \num{26.8}), lower p95 error on Hollywood JOB-Light (\num{12.7} vs. \num{46.7}), and higher p95 error on Hollywood JOB (\num{137} vs. \num{28.5}).

\topic{Learned Runtime Models}
Relative to the matched IMDb, \zeroshot has higher Hollywood medians on JOB-Light and JOB, but lower p95 on JOB and JOB-Complex (\num{66.9} vs. \num{130}, and \num{51.4} vs. \num{1080}). The signed boxes show that \zeroshot usually overpredicts Hollywood runtimes, although Hollywood JOB contains a severe underprediction. MSCN cost is mixed, with a larger Hollywood JOB-Light tail (\num{1287} vs. \num{68.8}) but a smaller Hollywood JOB-Complex tail (\num{83.2} vs. \num{228}).

\section{Conclusion \& Future Work}

With \system, we combine LLM semantic priors with deterministic temporal graph generation to produce IMDb/JOB-style benchmark instances with strict relational exports, adapted SQL, and executable labels. The generator exposes entity counts, company and keyword pools, role mixes, and yearly slate sizes as parameters for future scale factors and controlled workload variants.

In future work, we plan to introduce TPC-style scale factors and replay all existing learned cardinality estimators across these scaled IMDb-compatible datasets. We also aim to extend Hollywood's generation process to other schemas, such as those of TPC-H and TPC-DS, whose data generators rely on predefined statistical distributions rather than semantic generation.

\clearpage
\balance
\bibliographystyle{ACM-Reference-Format}
\bibliography{references}

\end{document}